\documentclass[fleqn,twoside]{article}%
\usepackage{amsmath}
\usepackage{amsfonts}
\usepackage{amssymb}
\usepackage{graphicx}%
\def\be{\begin{equation}}

\def\ee{\end{equation}}

\def\bi{\bibitem}

\begin{document}

\title {Revisiting Noether gauge symmetry for $F(R)$ theory of gravity}

\author{Nayem Sk.$^\dag$ and Abhik Kumar Sanyal$^{\S}$}
\maketitle

\noindent

\begin{center}
$\dag$Dept. of Physics, Ranitala High School, Murshidabad, India - 742135\\
$\S$Dept. of Physics, Jangipur College, Murshidabad, India - 742213\\

\end{center}

\footnotetext[1]{
Electronic address:\\

\noindent $\dag$nayemsk1981@gmail.com\\
\noindent $^{\S}$sanyal\_ ak@yahoo.com\\}


\begin{abstract}

\noindent Noether gauge symmetry for $F(R)$ theory of gravity has been explored recently. The fallacy is that, even after setting gauge to vanish, the form of $F(R) \propto R^n$ (where $n \ne 1$ is arbitrary) obtained in the process, has been claimed to be an outcome of gauge Noether symmetry. On the contrary, earlier works proved that any nonlinear form other than $F(R) \propto R^{\frac{3}{2}}$ is obscure. Here, we show that, setting gauge term zero, Noether equations are satisfied only for $n = 2$, which again does not satisfy the field equations. Thus, as noticed earlier, the only form that Noether symmetry admits is $F(R) \propto R^{\frac{3}{2}}$. Noether symmetry with non-zero gauge has also been studied explicitly here, to show that it does not produce anything new.

\end{abstract}

\maketitle

\section{Introduction}

$F(R)$ theory of gravity has been turned out to be very important to resolve dark energy issue that causes observed late time cosmic acceleration. It is well-known fact that any arbitrary $F(R)$ gravity is mathematically-equivalent to minimal scalar-tensor theory of gravity. This is achieved either by expressing the action $\int\sqrt{-g} d^4 x F(R)$ as $\int\sqrt{-g} d^4 x [F'(A)(R-A)+ F(A)]$ and making the scale transformation $g_{\mu\nu}\rightarrow e^{\phi}g_{\mu\nu}$ with $\phi = -\ln F'(A)$, or by expressing the action as $\int\sqrt{-g}d^4 x[R P(\phi) - Q(\phi)]$, whose variation with respect to $\phi$ yields $\phi = \phi(R)$, so that original action is regained. However, in the process, fourth order field equation reduces to second order ones. Study of such second order field equations following reconstruction procedure as well, has revealed many brilliant features of $F(R)$ theory of gravity (see, e.g., \cite{1}, for a recent review and the references therein). Of particular importance is the fact that \cite{1}, some versions of modified theories may be consistent with local tests and may provide a qualitatively reasonable unified description of inflation with the dark energy epoch. Nevertheless, modified gravity of specific form which describes acceptable accelerating universe (with realistic effective equation of state) is not physically equivalent to scalar-tensor theory \cite{2, 3}. Hence, the equivalent scalar-tensor gravity may not lead to accelerating FRW universe or, it may lead but with significantly different effective equation of state. There has also been several attempts to study fourth order theory of gravity, without invoking scalar-tensor equivalence \cite{4}. Since, it has been found that modification of gravitational action by including $R^{-1}$ term is unacceptable because such term would lead to a strong temporal instability resulting in a dramatic change of gravitational field of any gravitating body \cite{5}, therefore it is excluded. Thus, whatever might be the procedure, at the end of the day, one requires to choose some particular form of $F(R)$, suitable to explain recent cosmological observations. Since, symmetry has always played a dominant role in Physics, so invoking Noether symmetry in this connection is an elegant technique, because it provides a selection rule. Further,  Noether integrals are used to reduce the order of the field equations or even to solve them directly \cite{6}. \\

\noindent
Noether symmetry has been explored to find a form of $F(R)$ by several authors \cite{7, 8, 9, 10, 11, 12}, spanning the Lagrangian by a set of configuration space variables $(a, R, \dot a, \dot R)$. All attempts in this regard lead to $F(R) \propto R^{\frac{3}{2}}$, in the Robertson-Walker minisuperspace model, in vacuum and also in the matter dominated era. Such a form of $F(R)$ leads to inflation in the early Universe starting from initial radiation era. On the contrary, although it admits smooth transition from early deceleration to the late time accelerating phase in the matter dominated era, the decelerating phase tracks a solution in the form $a \propto t^{\frac{1}{2}}$, instead of the standard $t^{\frac{3}{2}}$. Further, radiation era yields $a \propto t^{\frac{3}{4}}$ against $t^{\frac{1}{2}}$ \cite{12}. Such solutions do not fit all the cosmic observations, particularly, these are grossly inconsistent with large scale structure formation (LSS) and WMAP data. Noether symmetry of $F(R)$ theory of gravity has also been explored in the presence of (minimally/non-minimally coupled) scalar field. The result is null, i.e., Noether symmetry remains obscure \cite{13}.\\

\noindent
Recently, some authors in the name of ``Noether Gauge Symmetry Approach in $F(R)$ Gravity" have claimed that other than $F(R) \propto R^{\frac{3}{2}}$, Noether symmetry in the presence of a gauge exists for $F(R) \propto R^n, ~ n \ne 1, \frac{3}{2}$, even after setting gauge to zero \cite{14}. This is crazy, since setting gauge to zero, gives back Noether symmetry without gauge and thus same old results should reappear. The reason behind the fallacy is that the authors \cite{14}, neither tried to satisfy all the Noether equations nor the field equations. Here we explore the same in section 3, and found that although Noether equations are satisfied only for $n = 2$,  the corresponding conserved current does not satisfy the field equations in general. Thus, indeed setting gauge to vanish the same old result $F(R) \propto R^{\frac{3}{2}}$ is admissible. In section 4, we perform Noether symmetry, with gauge, which has not been attempted earlier in metric formalism. The result is the same as before, i.e., only $F(R) \propto R^{\frac{3}{2}}$ is admissible. Thus, we conclude that inclusion of gauge in Noether symmetry formalism does not alter the situation. Before we proceed, in the following section we briefly review Noether gauge symmetry for a point Lagrangian.

\section{How gauge appears in Noether symmetry ?}

Among all the dynamical symmetries, transformations that map solutions of the equations of motion into solutions, one can single out Noether symmetries as the continuous transformations that leave the action invariant - except for boundary terms. In formal language, Noether symmetry for a point Lagrangian $L(q_i, \dot q_i, t)$ states that, ``For any regular system if there exists a vector field $X^{(1)}$, such that,

\be \left(\pounds _{X^{(1)}} + \frac{d\eta}{dt}\right)L = \left(X^{(1)} + \frac{d\eta}{dt}\right) L = \frac{d B}{d t},\ee

\noindent
in the presence of a gauge function $B(q_i, t)$, then there exists a conserved current,

\be I = \sum_{i}(\alpha_i - \eta \dot q_{i} )\frac{\partial L(q_{i}, \dot q_{i}, t)}{\partial \dot q_{i}} + \eta L(q_{i}, \dot q_{i}, t) - B(q_{i}, t) = \sum_{i}\alpha_i p_{i} - B(q_i, t) - \eta(q_i, t)H(q_i, p_i, t),\ee

\noindent
where,

\be X^{(1)} = X + \sum_{i}\Big[(\dot\alpha_{i} - \dot \eta\dot q_{i})\frac{\partial}{\partial\dot q_{i}}\Big],\ee

\noindent
is the first prolongation of $X$ given by,

\be X = \eta\frac{\partial}{\partial t} +\sum_{i} \alpha_i\frac{\partial}{\partial \dot q_{i}} ,\ee

\noindent
with, $\alpha_i = \alpha_i(q_i, t), ~\eta = \eta(q_i, t)$. It is apparent from the above expression of conserved current (2) that apart from the trivial Noether point symmetry $\frac{\partial}{\partial t}$, for which Noether integral is the Hamiltonian, different form of $\eta$, i.e., time translation considerably changes the non-trivial Noether current. Thus, one may obtain indefinitely large number of Noether integrals. The situation is different in the case of gravitational Lagrangian. Remember that conservation of the Hamiltonian is an outcome of the homogeneity of time, viz., ($\eta =$ constant), further since Hamiltonian here is constrained to vanish, Noether current reduces to

\be I =  \sum_{i}\alpha_i p_{i} - B(q_i, t).\ee

\noindent
Thus, $\eta$ does not play any additional significant role in conserved current for gravitational theories. Further, if one chooses the gauge to vanish ($B = 0$) at any stage, the conserved current remains the same as in the case of Noether symmetry without gauge and it is not possible to get extra (more than one) symmetry generators from the above definition. As already mentioned, vacuum $F(R)$ theory of gravity admits Noether symmetry only for $F(R) \propto R^{\frac{3}{2}}$. On the contrary, some authors \cite{14}, recently have claimed that the same system admits Noether symmetry for $F(R) \propto R^n$, where $n$ is arbitrary ($n \ne 0,~1$). Ridiculously, the authors \cite{14} claimed it to be the outcome of gauge Noether symmetry, even after setting gauge to vanish. Therefore it is important to
review the situation thoroughly.\\

\noindent
Before going into the issue discussed above, let us concentrate upon another important aspect in connection with gauge Noether symmetry of gravitational action. Although we are dealing with field theory, however to explore Noether symmetry, we reduce the Gravitational Lagrangian to point Lagrangian which does not contain time explicitly, while the gauge term $B$ might contain time explicitly. So, let us see how gauge term modifies Noether symmetry. For a point Lagrangian, Noether symmetry states that under infinitesimal transformations of co-ordinates ($q_i' = q_i + \epsilon \alpha_i(q_{i}, t)$) and time ($t'=t + \epsilon \eta(q_{i}, t)$), if the Hamilton's principal function remains unchanged then there exists a conserved current. Invariance of Hamilton's principal function under infinitesimal transformation implies,

\be \int L'(q_{i}', \dot q_{i}')dt' = \int L(q_{i}, \dot q_{i}) dt,\ee

\noindent
in the absence of explicit time dependance in the Lagrangian. Now, $L(q_{i}', \dot q_{i}', t')$ may be generated from $L'(q_{i}', \dot q_{i}')$ under the introduction of time dependent gauge in the following manner,

\be\int L(q_{i}',\dot q_{i}',t')dt'=\int L'(q_{i}',\dot q_{i}')dt'+\epsilon\int\frac{d B'(q_{i}', t')}{dt'}dt'.\ee

\noindent
From the above two equations (6) and (7) it follows that

\be\int L(q_{i}', \dot q_{i}', t') dt'= \int L(q_{i}, \dot q_{i}) dt + \epsilon\int \frac{d B'(q_{i}', t')}{dt'}dt'.\ee

\noindent
Now expanding $L(q_i', \dot q_i', t')$ and $B'(q_{i}', t')$ in Taylor series and under suitable transformations of coordinates, velocities and time, one ends up with,

\[ \int L(q_{i}, \dot q_{i}, t) dt - \int L(q_{i}, \dot q_{i}) dt 
+ \epsilon\frac{d}{dt}\int\left[\eta\left(L(q_i,\dot q_i, t) - \sum\dot q_i\frac{\partial L(q_i, \dot q_i, t)}{\partial \dot q_i}\right)\right]dt \]
\be + \epsilon\frac{d}{dt}\int\left[\sum\alpha_{i} \frac{\partial L(q_i, \dot q_i, t)}{\partial \dot q_i} - B(q_i, t)\right]dt = 0,\ee

\noindent
retaining only up to first order term. It is apparent from the above expression that, the first two terms does not get cancelled and so Noether integral as well as Noether symmetry remain obscure. The only way to cure it is to consider time independence of the Lagrangian, even after the introduction of gauge. This is possible if the gauge term $B$ is time independent as well. Hence, the very important point to remember about Noether gauge symmetry is that, for Lagrangian which does not contain time explicitly (as in the case of gravity), the gauge should have to be independent of time. Only under such condition equation (9) yields an integral of motion (2), which reduces to that given in (5) in the situation under consideration, where Hamiltonian is constrained to vanish.

\section{Noether gauge symmetry for $F(R)$ theory of gravity.}

With the understanding of the role of gauge term in the theory of gravity, let us review the work performed earlier \cite{14} in this connection. Although we have demonstrated that the gauge term should be time independent, still for the present purpose, we keep its time dependent form as it was considered by the earlier authors \cite{14} and show that Noether equations do not admit such time dependence in gauge. In spatially flat Robertson-Walker line element,

\be ds^2 = -dt^2+a^2\left[dr^2 + r^2 d\theta^2 + r^2\sin^2\theta d\phi^2\right],\ee

\noindent
the following action

\be A = \int d^4 x \sqrt{-g} ~F(R),\ee

\noindent
leads to a point Lagrangian

\be L = 6a \dot a^2 F' +6a^2\dot a\dot R F'' + a^3(F' R - F),\ee

\noindent
treating

\be R - 6\left(\frac{\ddot a + \dot a^2}{a^2}\right) = 0,\ee

\noindent
as a constraint of the theory and spanning the Lagrangian by a set configuration space variable $(a, R, \dot a, \dot R)$. The Noether gauge equation (1) is,

\[\alpha[6 \dot a^2 F^{\prime} + 12 a \dot{a}\dot R F^{\prime\prime} + 3a^2 (F^{\prime} R - F)]+ 
\beta[6 a \dot a^2 F^{\prime\prime} + 6a^2 \dot a \dot R F^{\prime\prime\prime} + a^3 F^{\prime\prime} R ]+ \]
\[\Big[\alpha_{,t}+(\alpha_{,a} - \eta_{,t})\dot a\Big](12a\dot a F^{\prime}+ 6a^2 \dot R F^{\prime\prime})+ 
\Big[\alpha^{\prime}\dot R - \dot a^2\eta_{,a} -\dot a\dot R \eta^{\prime}\Big](12a\dot a F^{\prime}+ 6a^2 \dot R F^{\prime\prime})+ \]
\[\Big[\beta_{,t} + \beta_{,a}\dot a+(\beta^{\prime}-\eta_{,t})\dot R - \dot R^2\eta^{\prime} - \dot a\dot R \eta_{,a}\Big](6 a^2\dot a F^{\prime\prime})+ \Big[\eta_{,t} + \dot a\eta_{,a} + \dot R\eta^{\prime}\Big]\Big[6 a \dot a^2 F^{\prime} + 6a^2\dot a\dot R F^{\prime\prime}+a^3(F^{\prime}R - F)\Big] \]
\be = \frac{d B}{d t} = B_{,t} + \dot a B_{,a} + \dot R B^{\prime}.\ee

\noindent
Equating coefficients to zero as usual, one obtains the following set of Noether equations

\be a F^{\prime}\eta_{,a} = 0  \ee
\be 6a^2 F^{\prime\prime}\eta' = 0 \ee
\be \alpha F' + a\beta F''+2a\alpha_{,a} F'+a^2 \beta_{,a} F'' - aF'\eta_{,t} = 0 \ee
\be 12 aF'\alpha_{,t} + 6a^2 \beta_{,t}F'' = B_{,a} \ee
\be 6a^2 F''\alpha' = 0  \ee
\be 6a^2 F''\alpha_{,t} = B' \ee
\be 2a\alpha F'' + a^2 \beta F'''+a^2 F''(\alpha_{,a}+\beta'-\eta_{,t})+2aF'\alpha' = 0  \ee
\be a^2(3\alpha+a\eta_{,t})(F' R - F)+a^3 F'' R\beta = B_{,t}. \ee

\noindent
where, dash($'$) represents derivative with respect to $R$. These set of equations clearly depict that for a nonlinear form of $F(R)$ (i.e., $F^{\prime\prime} \ne 0$), $\eta = \eta (t), ~\alpha = \alpha(a)$ and $\beta = \beta(a, R)$, provided the gauge is set to zero $(B = 0)$ as was assumed in reference \cite{14}. Hence, equations (15, 16, 18 - 20) are trivially satisfied and rest of Noether equations, viz., (17, 21 and 22)  now reduce to,

\be (\alpha +2a\alpha_{,a})F' + a(\beta + a \beta_{,a}) F'' = aF'\eta_{,t}  \ee
\be (2\alpha + a\alpha_{,a} + a\beta')F'' + a\beta F''' = a F''\eta_{,t} \ee
\be 3(R F' - F)\alpha+\beta a R F'' = -a(R F' - F)\eta_{,t}. \ee

\noindent
Clearly, $\eta$ has to be linear in time. The above set of equations is still different from the ones obtained earlier
\cite{7, 8, 9, 10, 11, 12}, since here time translation is involved. In general, the above set of equations admit a solution in the following form,

\be \eta = c_1 t + c_2,\; F = f_0 R^{\frac{3}{2}},\; \alpha = \frac{2}{3}c_1 a - \frac{c_3}{3 a},\; \beta= -2R\left[c_1 - \frac{c_3}{3 a^2}\right],\;\;I =  2c_1a^4\frac{d}{dt}\left(\frac{\sqrt R}{a}\right) - c_3\frac{d}{dt} (a\sqrt R),\ee
It is interesting to note that above integral of motion does not satisfy the field equations corresponding to $F(R) = f_0 R^{\frac{3}{2}}$ other than for $c_1 = 0$. The obvious reason (explained in section 2) behind this fact is that the conservation of Hamiltonian requires homogeneity of time, i.e., $\eta = $ constant. However under such condition $\eta$ turns out to be time independent and the above conserved current reduces automatically to the one [$\frac{d}{dt}(a\sqrt R) = $ constant] obtained earlier \cite{7, 8, 9, 10, 11, 12}. However, \cite{14} claimed in addition that, under the choice,

\[F(R) = f_0 R^n,~~n \ne 0, 1, \frac{3}{2},\]

\noindent
$f_0$ being a constant, the above set of equations admit a solution in the form

\be\eta = c_1 t + c_2,\;\; \alpha(a) = \frac{c_3 - c_1}{3}a,\;\;\beta = -2c_1 R.\ee

\noindent
On the contrary, what we observe is that, if one restricts $\beta \ne \beta (a)$ a-priori, then $n (\ne \frac{3}{2})$ apparently becomes arbitrary. But then, it is not difficult to check that all the three Noether equations (23) through (25) are now satisfied only for $n = 2$ and the corresponding conserved current takes the form
\be a^3 \dot R =I.\ee

\noindent
Now the field equations for $F(R) = f_0 R^2$ are,

\be \ddot R + 2\frac{\dot a}{a}\dot R + R\left(2\frac{\ddot a}{a} + \frac{\dot a^2}{a^2} -\frac{R}{4}\right) = 0,\ee

\be \frac{\dot a}{a}\dot R + R\left(\frac{\dot a^2}{a^2} -  \frac{R}{12}\right) = 0.\ee

\noindent
Noether current is not an independent equation, it is just the first integral of some combination of the field equations and so one has to check if it satisfies the field equations. It is obvious that the field equations will not be satisfied, unless $\eta =$ constant, as mentioned earlier. One can still check in a straight forward manner, that the conserved current obtained here, does not satisfy the field equations in general, other than for the trivial situation, viz., $R = 0$ which makes $c_1 = 0$, making $\eta =$ constant. Thus, restricting to $\beta \ne \beta(a)$ is not allowed and hence, the only form of $F(R)$ that admits Noether symmetry is $F(R) \propto R^{\frac{3}{2}}$, if one sets the gauge term to vanish. Hence, the result obtained by \cite{14} is not correct.\\

\section{General solution with Nonzero gauge.}

In the previous section we have proved that Noether symmetry with zero gauge yields nothing other than $F(R) \propto R^{\frac{3}{2}}$. In this section, we proceed to see what a non-zero gauge term contributes, which has not been studied earlier. So, let us try to find a solution in the form $F(R) = f_0 R^{n}$, keeping a non-vanishing gauge term in the Noether equations. Thus in view of equations (15) and (16) one finds, $\eta$ is independent of both $a$ and $R$, i.e., $\eta = \eta(t)$, while (19) depicts $\alpha$ is independent of $R$, i.e., $\alpha = \alpha(a, t)$. Equation (17) thus implies $\beta$ must be linear in $R$. In view of which equations (18) and (20) imply, $B \propto R^{n-1}$. Equation (22) is now satisfied provided $B$ is independent of $t$, i.e., $B = B(a, R)$. Note that in general, we have proved this fact in section (2) that for Noether symmetry to exists the gauge has to be independent of time, if the Lagrangian of the system does not contain time explicitly. Since $B \ne B(t)$, as a result, equation (20) implies, $\alpha$ must be linear in $t$. This further indicates in view of equation (18) that $\beta$ should also be linear in $t$, while $\eta$ has to be quadratic in $t$ as is observed from equations (17) and (22). Hence finally we have,

\be \eta = c_1 t^2 + c_2,\;\;\;\alpha = t\alpha_1(a),\;\;\;\beta = t  \beta_1(a) R,\;\;\; B = F_0 R^{n-1}B_1(a).\ee

\noindent
Equation (20) now indicates \be B_1 = 6n a^2\alpha_1,\ee

\noindent
while, using equation (32) in (18) one finds,

\be \beta_1 = \frac{\alpha_{1,a}}{n-1}.\ee

\noindent
Using the above equation, viz., (33) in equations (17), (21) and (22), following three linear differential equations in $\alpha$ are obtained, which are to be solved simultaneously,

\be a^2 \alpha_{1,aa} + 3 a \alpha_{1,a} + \alpha_1 = 2c_1 a.\ee
\be a\alpha_{1,a} + \alpha_1 = c_1 a.\ee
\be a \alpha_{1,a} +3\Big(\frac{n-1}{n}\Big) \alpha_1 = -2\Big(\frac{n-1}{n}\Big) c_1 a.\ee

\noindent
These three equations are satisfied simultaneously for

\be \alpha_1 = \frac{c_3}{a},\;\;c_1 = 0,\;\;and\;\; n = \frac{3}{2}.\ee

\noindent
Hence, $\eta = c_2$ becomes a constant and it appears that we arrive at our same old result. But this is not true, since now we have

\be F = F_0R^{\frac{3}{2}},\;\; \eta = c_2,\;\;\alpha = c_3\frac{t}{a},\;\;\beta =
-2 c_3\frac{Rt}{a^2},\;\;B = 9 F_0 c_3 a\sqrt R,\ee as a result of
which the conserved current is

\be I = \left[\frac{d}{dt}(a\sqrt R)t - a\sqrt R\right].\ee

\noindent
Above integral of motion may at once be integrated at once to yield,

\be a\sqrt R = d_0 t - I,\ee
where, $d_0$ is an integration constant. It is fairly interesting to note that the above equation is identical in disguise, to the one $\frac{d}{dt}(a\sqrt R) = I$ obtained earlier without gauge \cite{7, 8, 9, 10, 11, 12}. Thus the solution also is the same

\be a = \sqrt{\frac{1}{72}\left[d_0^2 t^4 - 4I d_0 t^3 + 12 I^2 t^2 + 12d_1 t  + 12 d_2\right]},\ee
where, $d_1, \;d_2$ are constants of integration. So, the same old solution reappears as soon as $c_1 = 0$ i.e., $\eta =$ constant. Thus we observe that inclusion of gauge term does not yield anything new.

\subsection{Case II}

Thus finally we are left with one important case to study, i.e.,
setting $\beta \ne \beta(a)$ a-priori. In this case we have,

\be \eta = c_1 t^2 + c_2,\;\;\alpha = \alpha_1(a)t,\;\; \beta = \beta_0
Rt,\;\;and\;\;B = B_{1}(a) R^{(n-1)}\ee Equations (17) and (21)
then take the following forms

\be 2a\alpha_{1,a} + \alpha_1 = [2c_1 -(n-1)\beta_0]a,\ee
\be a\alpha_{1,a} + 2\alpha_1 = [2c_1 -(n-1)\beta_0]a,\ee
while equation (22) gets solved to yield
\be\alpha_1 = -\left(\frac{2c_1+n\beta_0}{3}\right)a.\ee
This solution satisfies the above pair of differential equations
under the condition

\be\beta_0 = -4c_1.\ee
Finally equation (18) gets solved to yield

\be B_{1} = -\frac{8F_0 n(n-2)c_1}{3}a^3 + c_3.\ee
Equation (20) then demands that the integration constant $c_3$ has
to vanish and $n = \frac{7}{8}$. This result was found earlier \cite{15} and was rejected since it does not yield analytical solution. On the contrary we find that the conserved current

\be I = \left[\left(3 a^2 \dot a R^{-\frac{1}{8}}-\frac{1}{8}a^3\dot R R^{-\frac{9}{8}}\right)t - a^3 R^{-\frac{1}{8}}\right] 
= \frac{d}{dt}\left(a^3 R^{-\frac{1}{8}}\right)t - a^3 R^{-\frac{1}{8}}\ee

\noindent
which is integrated to yield $a^3R^{-\frac{1}{8}} = e_0 t$, and may be expressed as

\be\frac{d}{dt}\left(a^3 R^{-\frac{1}{8}}\right) = e_0,\ee
where, $e_0 = t_0 - I$, $t_0$ being the constant of integration. The field equations for $F(R) = R^{\frac{7}{8}}$ are the following,

\be 2\frac{\ddot a}{a} + \frac{\dot a^2}{a^2} - \frac{\ddot R}{8 R} -\frac{\dot a\dot R}{4 a R} + \frac{9}{64}\frac{\dot R^2}{R^2} + \frac{R}{14} = 0.\ee

\be 3\frac{\dot a^2}{a^2} - \frac{3\dot a\dot R}{ 8 a R} + \frac{R}{14} = 0.\ee

\noindent
It may be checked through a little complicated algebra that the conserved current does not satisfy the above field equations. Although we have made a detailed analysis, nevertheless, from our early understanding it is known a-priori that Noether symmetry will be obscure in this case, due to the presence of explicit time dependence in $\eta$. If one makes $\eta$ constant, by setting $c_1 = 0$, $\beta_0$ vanishes as well and so both the gauge term and $\beta$ vanish. We already know that in the absence of gauge term symmetry yields $F(R) \propto R^{\frac{3}{2}}$, keeping $\beta$ arbitrary. Hence, one can not expect anything new by constraining $(\beta \ne a)$. So, this case is eliminated. \\

\subsection{Case III}

Let us set $B \ne B(a)$ a-priori, but then unless $B \ne B(t)$, equation (22) is not satisfied, so $B = B(R)$. This gauge condition amounts in $\alpha(t) \propto t$ in view of equation (20), so that $\alpha = \alpha_0\frac{t}{a^2}$ and $B = B_0 R^{n-1}$. Equation (18) implies $\beta(t) = \frac{2\alpha_0 \beta _0}{(1-n)a^3} Rt$, in view of equations (17) and (21) $\eta(t) = c_1 t^2 + c_2$. However, equations (17), (21) and (22) are not satisfied. The obvious reason that $\eta$ has explicit time dependance has been mentioned earlier. Thus this case is abandoned.

\subsection{Case IV}

Finally, if one starts with the fact revealed in earlier cases, viz., $\alpha \ne \alpha(t),\;\beta \ne \beta(t),\;B \ne B(t)$, thus equations (15) through (22) finally yield,

\be \eta = \eta(t),\;\;\alpha = \alpha(a),\;\;\beta = \beta(a,
R),\;\; B = B_0\;\;-\;a\;constant.\ee

\be \frac{\alpha}{a} + 2 \alpha_{,a} + (\beta +
a\beta_{,a})\frac{F''}{F'} = \eta_{,t}.\ee

\be 2\frac{\alpha}{a} +  \alpha_{,a} + \beta \frac{F'''}{F''}+
\beta' = \eta_{,t}.\ee

\be 3\frac{\alpha}{a} + \frac{F''}{F' R - F}R\beta =
-\eta_{,t}.\ee

\noindent
These set of equations lead to our old result, viz., $F(R) = F_0 R^{\frac{3}{2}}$, while the constant gauge does not make any difference and can be set to zero, without loss of generality. The conserved current is the one obtained in equation (26), which does not satisfy the field equation unless $c_1 = 0$, i.e., $\eta =$ constant and the same old result reappears.

\section{Concluding remarks.}
Noether symmetry for $F(R)$ theory of gravity for spatially flat or non-flat Robertson-Walker space-time in vacuum or in the presence of matter in the form of dust yields nothing other than $F(R) \propto R^{\frac{3}{2}}$. Such a form is quite good to explain early Universe, since there exists a smooth transition from radiation type of solution $a \propto \sqrt t$ to power law inflation. However, it is not suitable for late stage of cosmological evolution, since it yields $a \propto t^{\frac{4}{3}}$ in the radiation era and early deceleration in the matter dominate era evolves as $a \propto \sqrt t$. These results put up sever problem in explaining structure formation and observed CMBR data.\\

\noindent
Recently some authors have claimed that gauge Noether symmetry for $F(R)$, in spatially flat metric yields arbitrary powers of $R$, i.e., for $F(R) \propto R^n$ even after setting gauge to vanish. Such a result was obtained simply because the authors did not check all the Noether equations. We have shown that Noether equations are satisfied for $n = 2$, but the corresponding conserved current again does not satisfy the field equations. Thus any form of $F(R)$ other than $R^{\frac{3}{2}}$ is obscure.\\

\noindent
Indeed it is interesting to check if a non-vanishing gauge term yields anything else, which has not been attempted earlier for pure gravity. We have found that the same old result, ie., $F(R) \propto R^{\frac{3}{2}}$ emerges. The conserved current appears to be different at first sight. However, it has been shown to be the same old one in disguise. Hence, it yields the same cosmological solution (41).\\

\noindent
We have not included matter term here, just to compare results obtained in \cite{14}. Nevertheless, we have also checked that inclusion of a pressureless dust term in the form $A_m = -\int\sqrt{-g} d^4x\rho_{m0} a^{-3}$ in the action (11), or equivalently the present matter-density $- \rho_{m0}$ in the point Lagrangian (12) does not alter the situation.\\

\noindent
A couple of important points has been elaborated. These are (i) gauge term has to be time independent for time independent Lagrangian and (ii) for the situation under consideration, where, Hamiltonian is conserved and is constrained to vanish, one has to impose homogeneity of time by setting $\eta =$ constant.\\

\noindent
Thus, finally we conclude that Noether symmetry does not yield anything new in the presence of gauge, other than making things more complicated.

\end{document}